\documentclass[prl, twocolumn, 10pt, amsmath, amssymb, showpacs]{revtex4}
\usepackage{amssymb, stmaryrd, graphicx, epsfig}

\newcommand{\diag}{\mathop{\rm diag}\nolimits}

\def\be{\begin{eqnarray}}
\def\ee{\end{eqnarray}}
\def\bee{\begin{eqnarray*}}
\def\eee{\end{eqnarray*}}

\begin{document}
\title{Universal resources for measurement--based quantum computation} 
\author{Maarten Van den Nest$^1$, Akimasa Miyake$^{1,2}$, Wolfgang D\"ur$^{1,2}$ and Hans J. Briegel$^{1,2}$}

\affiliation{$^1$ Institut f\"ur Quantenoptik und Quanteninformation der \"Osterreichischen Akademie der Wissenschaften, Innsbruck, Austria\\
$^2$ Institut f{\"u}r Theoretische Physik, Universit{\"a}t Innsbruck,
Technikerstra{\ss}e 25, A-6020 Innsbruck, Austria
}
\date{October 13, 2006}

\def\makeheadbox{}

\begin{abstract}

We investigate which entanglement resources allow universal measurement--based quantum computation via single--qubit operations. We find that any entanglement feature exhibited by the 2D cluster state must also be present in any other universal resource. We obtain a powerful criterion to assess universality of graph states, by introducing an entanglement measure which necessarily grows unboundedly with the system size for all universal resource states. Furthermore, we prove that graph states associated with 2D lattices such as the hexagonal and triangular lattice are universal, and obtain the first example of a universal non--graph state. 

\end{abstract}


\maketitle

{\bf Introduction.---} Quantum computation is a promising attempt to utilize the laws of quantum physics for novel applications. Indeed, it was shown that problems such as factoring or database search can be performed much faster on a quantum computer than on any known classical device. Despite of these exciting perspectives, the question: \emph{what are the essential resources that give quantum computers their additional power over classical devices?} is still poorly understood. Various models for a quantum computer exist, each based on different concepts, which indicates that there may not be a straightforward answer to this difficult question. The new paradigm of {\em measurement--based quantum computation} (MQC) \cite{Ra01,Go99,Ni03, Pe04}, with the one--way quantum computer \cite{Ra01} and the teleportation--based model \cite{Go99,Ni03} as the most prominent examples, has lead to a new and fresh perspective in these respects. In particular, these and other studies \cite{Jo03} highlight the central role of entanglement in quantum computation.

In MQC, quantum information stored in a quantum state is processed by performing sequences of adaptive measurements. This is in striking contrast to the quantum circuit model, where unitary operations are realized via coherent evolution. While the teleportation--based models \cite{Go99,Ni03} use joint (i.e. entangling) measurements on two or more qubits, thereby performing sequences of teleportation--based gates, the one--way model \cite{Ra01} uses a highly entangled state, the \emph{cluster state} \cite{Br01}, as a universal resource which is processed by single-qubit measurements. Unified descriptions of all measurement--based models have recently been proposed in Refs.~\cite{Ve04,Ch05}. 

Here we focus on the one--way  model, where the resource character of entanglement is particularly highlighted, as it is clearly separated from the processing via {\em local} measurements which do not act as additional source for entanglement. The distinct features of the one-way model also allow us to cast the introductory question into a much more concise form, viz. \emph{what are the essential properties of the cluster state that make it a universal resource?} In this letter we will investigate this question. The main objectives of our investigation are (i) to understand which states, other than the cluster states, are universal resources for MQC and (ii) to gain insight in the role of entanglement in this matter. It is believed that the high degree of entanglement in the (2D and 3D) cluster states plays an important role in the universality of these states, but the explicit entanglement features accounting for this have not been identified yet. For  example, it has recently been found that one--way quantum computations implemented on certain graph states (such as the 1D cluster states and the 
Greenberger-Horne-Zeilinger (GHZ) states) can be simulated efficiently on a classical computer \cite{Ni05}. Nevertheless, similar to the 2D and 3D cluster states, these states are highly entangled, e.g. in that they maximally violate certain Bell inequalities \cite{Gu04}.

In our study we will slightly extend the framework of the one--way model by allowing \emph{arbitrary} local operations and classical communication (LOCC) to be implemented on resource states, rather than restricting ourselves to local measurements, hence emphasizing the role of entanglement in this context. Within this general framework we obtain two main results. First we find large classes of states, including various families of graph states \cite{He06}, that are not universal, by identifying an entanglement measure that needs to grow unboundedly with the number of qubits for all universal resources. Second, we provide new examples of universal resource states, including graph states corresponding to hexagonal, triangular and Kagome lattices, as well as an example of a universal non--graph state.

{\bf Definition of universal resources.---} Let us briefly recall the general procedure of the one--way model endowed with a 2D cluster state as a universal resource \cite{Ra01}. It is capable to simulate any unitary evolution $U$, 
acting on a standard input state $|+\rangle^n$ (the $n$-qubit product state of
a $+1$ eigen state of $\sigma_x$), and to produce a corresponding 
output state $|\phi\rangle := U |+\rangle^n$ deterministically, as follows. 
(i) A 2D cluster state $|C_{k\times k}\rangle$ is prepared, which is a particular instance of a \emph{graph state}. A graph state on $m$ qubits is the joint eigenstate of $m$ commuting correlation operators $K_a:= \sigma_x^{(a)}\bigotimes_{b\in N(a)} \sigma_z^{(b)}$, where $N(a)$ denotes the set of neighbors of qubit $a$ in the graph \cite{He06}. The 2D cluster state is obtained if the underlying graph is a $k\times k$ square lattice (thus $m=k^2$).  
(ii)  A sequence of adaptive one--qubit measurements is
implemented on some subset of qubits in the cluster. (iii) After these
measurements, the state of the system has the form
$|\xi^{\alpha}\rangle |\phi^{\alpha}_{\mbox{\scriptsize out}}\rangle$,
where $|\phi^{\alpha}_{\mbox{\scriptsize out}}\rangle= \Sigma^{\alpha}U|+\rangle^n$ is the desired output state up to a multi--qubit Pauli operator $\Sigma^{\alpha}$ which depends on the measurement outcomes $\alpha$ obtained in (ii), and the measured qubits are in a product state $|\xi^{\alpha}\rangle$ which also depends on the measurement outcomes $\alpha$. Note that the required size of the cluster and the choice of local measurements are determined by $U$.

Having the above procedure in mind, we will propose a definition of a \emph{universal resource} for MQC. Before doing this, two remarks are in order. 
First, universality is a property that will be
attributed, not to a single state, but to a
set of infinitely many states $\Psi =\{|\psi_1\rangle,
|\psi_2\rangle, \dots\}$ \cite{Foot}. 
When considering the 2D cluster state model, it is
indeed clear that it is not one cluster state which forms
a universal resource, but rather the family of all
2D cluster states. This is most evident in step (i) above,
where the size of the cluster state depends on the unitary
to be simulated. Second, here we are only interested in \emph{universality} of a family of states $\Psi$ and
not necessarily 
in its \emph{efficiency} as a resource for quantum computation. That is, in our
definition we will only require that it is possible, in
principle, to simulate any unitary operation $U$ by implementing LOCC on a suitable state
$|\psi\rangle\in\Psi$, dependent on $U$, and we will not
consider the size (support) of $|\psi\rangle$ relative to the complexity of $U$.

We are now ready to formulate the following definition. A family $\Psi$ is called a
\emph{universal resource for MQC} if for each  state
$|\phi\rangle$ on $n$ qubits there exists a state
$|\psi\rangle \in \Psi$ on $m$ qubits, with $m\geq
n$, such that the transformation $|\psi\rangle \mapsto
|\phi\rangle|+\rangle^{m-n}$
 is possible deterministically (with probability 1) by LOCC,
denoted symbolically by $|\psi\rangle \geq_{\mbox{\tiny LOCC}} |\phi\rangle$.
That is, using only states within the family $\Psi$ as resource,
any state $|\phi\rangle$ can be prepared, and equivalently 
any unitary operation $U$ acting on an input state $|+\rangle^{n}$,
given as $|\phi\rangle= U|+\rangle^{n}$, can be implemented. 
This definition is in the spirit of the one--way model.

The following elementary observation immediately follows from the universality 
of the 2D cluster states. 

{\bf Observation 1:} {\it A set of states $\Psi$ is a universal resource for MQC if and only if all 2D cluster states $|C_{k\times k}\rangle$ (for all $k$) can be prepared from the set $\Psi$ by LOCC.}

The above insight, while indeed simple, leads to powerful techniques to both obtain no--go theorems, providing examples of sets $\Psi$ which are \emph{not} universal resources, and to construct several nontrivial examples of universal resources, other than the sets of 2D or 3D cluster states. 

{\bf Non--universality and entanglement.---} First we study no--go results. Here, our general strategy will be the following. Let $\Psi$ be a given set of states, of which one wishes to assess whether it is a universal resource. To do so, suppose one can identify a functional $E(|\psi\rangle)$ exhibiting the following two properties:
\begin{itemize}
\item[(P1)] $E(|\phi\rangle)\geq E(|\phi'\rangle)$ whenever $|\phi\rangle\geq_{\mbox{\tiny LOCC}}|\phi'\rangle$,
\item[(P2)] $\sup_{\forall |\phi\rangle} E(|\phi\rangle) > \sup_{|\psi\rangle\in\Psi} E(|\psi\rangle)$. 
\end{itemize} 
Property (P1) states that the measure $E$ cannot increase under LOCC, and (P2) states that the supremal value of $E$, when the supremum is taken over all states, is not reached on the family $\Psi$. Clearly, (P1) and (P2) imply that the set $\Psi$ cannot be a universal resource. 
Note that from the universality of the 2D cluster states one has
$\sup_{\forall |\phi\rangle} E(|\phi\rangle) = \sup_k E(|C_{k\times k}\rangle)$.
Using this property as a convenient reference, suitable choices for the measure $E$ will give rise to examples of non--universal resources. 

As the requirements (P1) and (P2) are in fact quite general, a priori there exist several candidates for measures $E$. As our main example we focus on a measure having its roots in graph theory, and which will prove particularly useful to assess whether sets of \emph{graph states} are universal resources. This measure will be called \emph{(entropic) entanglement width}, as its definition is a direct generalization of a graph invariant called \emph{rank width} \cite{Oum}. The entanglement width of an $m$-qubit state $|\psi\rangle$ is defined via the minimization of the bipartite entanglement entropy of
$|\psi\rangle$ over a specific class of bipartite splits, as follows. 
First, let $T$ be a subcubic tree graph, i.e. a connected graph without cycles and every node in the tree is incident with at most three edges. The nodes with one edge are called the leaves of $T$. We consider trees $T$ with exactly $m$ leaves, which are identified with the set of qubits $V:=\{1, \dots, m\}$. For any edge $e=\{i, j\}$ of $T$, let $T\setminus e$ be the graph
obtained by deleting the edge $e$ from $T$. The graph 
$T\setminus e$ then consists of exactly two connected
components, which induce a bipartition
$(A_{T}^e, B_{T}^e)$ of the set of qubits $V$.
Next, define $\alpha_T(|\psi\rangle)$ to be the maximum, over all edges $e$ of $T$, of the quantity $E_{A_{T}^ e}(|\psi\rangle)$, where the latter is the bipartite entanglement entropy of $|\psi\rangle$ with respect to the bipartition $(A_{T}^e, B_{T}^e)$. The entanglement width of $|\psi\rangle$ is now defined as
\be E_{\mbox{\scriptsize{wd}}}(|\psi\rangle):= \min_T\ \alpha_T(|\psi\rangle)=\min_T\ \max_{e\in T}\ E_{A_{T}^ e}(|\psi\rangle),\ee where the minimization is taken over all subcubic trees $T$ with $m$ labeled leaves. 

An exact evaluation of entanglement width in a generic state is likely to be a hard problem, given the min--max problem in the definition. However, the strength of this measure lies in its connection to the graph--theoretical measure \emph{rank width} ${\rm rwd}(G)$ \cite{Oum}, where good upper and lower bounds -- that are sufficient for our purpose -- are known. The entanglement width of a graph state coincides with the rank width of the underlying graph, which follows from the equivalence of the cut--rank \cite{Oum} of an adjacency matrix and the bipartite entanglement of the corresponding graph state \cite{He06}.

Note that entanglement width is invariant under local unitary operations and that it vanishes on complete product states. Furthermore, it is non-increasing under LOCC operations.  To see this, let $|\psi\rangle$ be an $m$-qubit state which is convertible by LOCC into another $m$-qubit state  $|\psi'\rangle$. We show that $E_{\mbox{\scriptsize{wd}}}(|\psi\rangle)\geq E_{\mbox{\scriptsize{wd}}}(|\psi'\rangle)$. Let $T_0$ be a subcubic tree such that  $\alpha_{T_0}(|\psi\rangle)  = E_{\mbox{\scriptsize{wd}}}(|\psi\rangle)$ and  let $e_0$ by an edge of $T_0$ such that $E_{A_{T_0}^{e_0}}(|\psi'\rangle) = \alpha_{T_0}(|\psi'\rangle).$  We then have
\be E_{\mbox{\scriptsize{wd}}}(|\psi\rangle)&=&\alpha_{T_0}(|\psi\rangle)\geq E_{A_{T_0}^{e_0}}(|\psi\rangle) \nonumber\\ &\geq& E_{A_{T_0}^{e_0}}(|\psi'\rangle) = \alpha_{T_0}(|\psi'\rangle)\geq E_{\mbox{\scriptsize{wd}}}(|\psi'\rangle).
\ee  
Using this result, we will show that entanglement width satisfies property (P1).
Let $|\psi\rangle$ and $|\phi\rangle$ be states on $m$ and $n$ qubits ($m\geq n$) respectively, such that $|\psi\rangle\geq_{\mbox{\tiny LOCC}}|\phi\rangle$. Equivalently, $|\psi\rangle$ can be converted into $|\phi\rangle|+\rangle^{m-n}$ by means of LOCC and thus $E_{\mbox{\scriptsize{wd}}}(|\psi\rangle)\geq E_{\mbox{\scriptsize{wd}}}(|\phi\rangle|+\rangle^{m-n}).$ As the states $|\phi\rangle|+\rangle^{m-n}$ and $|\phi\rangle$ have equal entanglement width, one finds that $E_{\mbox{\scriptsize{wd}}}(|\psi\rangle)\geq$ $E_{\mbox{\scriptsize{wd}}}(|\phi\rangle)$. 

Regarding (P2), we show that $E_{\mbox{\scriptsize{wd}}}(|C_{k\times k}\rangle) \geq 
\log_2(k+2)-1$ using a graph invariant called
\emph{clique width} $\mbox{cwd}(G)$, since ${\rm cwd}(C_{k\times k}) = k+1$ if 
$k\geq 3$ \cite{Go00} and $\mbox{rwd} \geq \log_2 (\mbox{cwd}+1)-1 $ 
\cite{Oum}.
Thus, $E_{\mbox{\scriptsize{wd}}}(|C_{k\times k}\rangle)$ diverges when $k$ tends to infinity.  
This leads to the following result.

{\bf Theorem 1:} {\it Any universal resource for MQC must have an unbounded  entanglement width.}

 This result allows us to rule out several classes of graph states as being non--universal resources, namely all classes having a bounded rank width. The list of non--universal graph states includes (the reader is referred to the literature for
definitions) {\em cycle graphs, cographs, graphs locally equivalent to trees, graphs of bounded tree width, graphs of bounded clique width or distance--hereditary graphs}. 
In particular, 1D cluster states $|C_k\rangle$ are not universal, since $E_{\mbox{\scriptsize{wd}}}(|C_k\rangle)= 1$ for every $k$ \cite{dist_her}. More generally, all graph states with bounded tree width twd$(G)$ are not universal \cite{twd}, which follows from the inequality $E_{\mbox{\scriptsize{wd}}}(|G\rangle) \leq 4\cdot 2^{\mbox{\scriptsize twd}(G) -1}+1$ \cite{cwd}. This also implies that the family of GHZ states (which correspond to tree graphs) is not a universal resource. These results support recent findings that any one--way computation performed on 1D cluster states or graph states with small tree width can efficiently be simulated on a classical computer \cite{Ni05}.

As a second example of a measure satisfying (P1) and (P2), consider the \emph{localizable entanglement} $E^L_{ab}(|\psi\rangle)$ (of an arbitrary state $|\psi\rangle$) between pairs of qubits $a$ and $b$ measured by the concurrence \cite{Ve_loc04}. This quantity is an entanglement monotone for $2 \times 2 \times l$ systems \cite{Go05} and also fulfills property (P1). As $E^L_{ab}(|C_{k\times k}\rangle)=1$ for every pair of qubits $a$ and $b$, deterministic generation of cluster states by means of LOCC requires as a necessary (but by no means sufficient) condition that there exists at least one pair of qubits in the system having unit localizable entanglement. This simple condition already identifies numerous non--universal resources, such as e.g. the family of $W$--states. A stronger condition can be obtained by considering the maximal size $N_{\mbox{\scriptsize LE}}(|\psi\rangle)$ of a subset of qubits in the system  in which all pairs of qubits have unit localizable entanglement. As $N_{\mbox{\scriptsize LE}}(|\psi\rangle)$ fulfills (P1) and $N_{\mbox{\scriptsize LE}}(|C_{k\times k}\rangle)=k^2$, it follows that the measure $N_{\mbox{\scriptsize LE}}(|\psi\rangle)$ must grow unboundedly on any universal resource. In particular, this implies that any class of states associated with some geometry for which the localizable entanglement $E^L_{ab}$ exhibits a decay with the distance $\|{\bf x}_a - {\bf x}_b\|$, cannot be universal resource. E.g., ground states of strongly correlated spin systems on any type of lattice where the above decay of $E^L_{ab}$ is observed, are not universal.  Notice that even a diverging entanglement length is not sufficient to guarantee universality. 

\begin{figure}[b]
\begin{minipage}{4.25cm}
\includegraphics[width=4.25cm,clip]{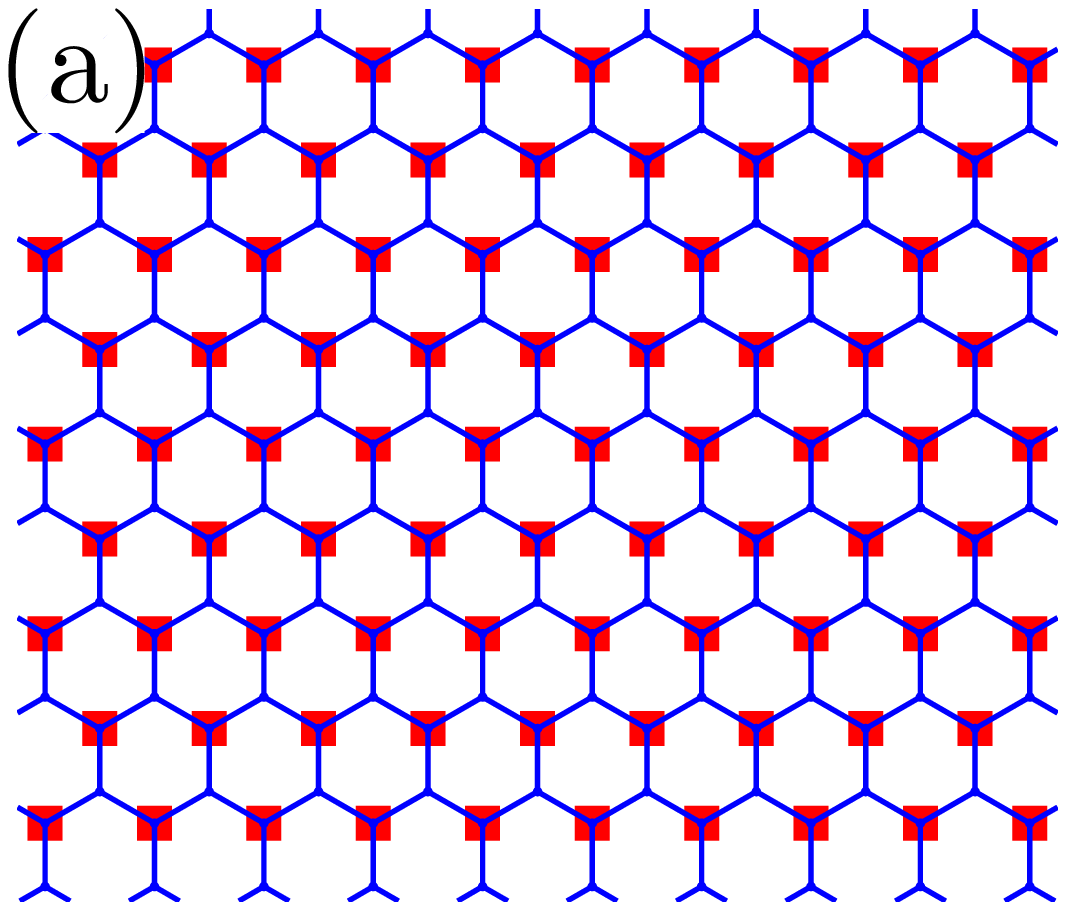}
\end{minipage}
\begin{minipage}{4.25cm}
\includegraphics[width=4.25cm,clip]{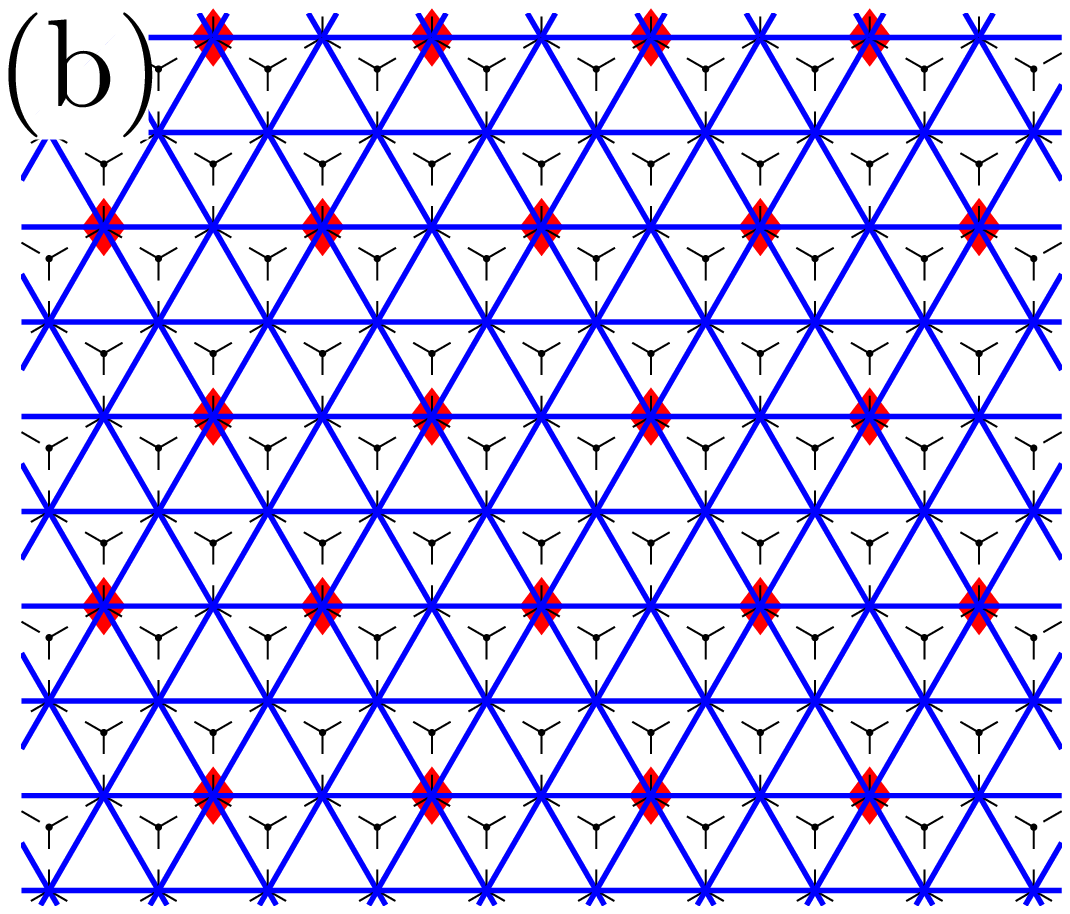}
\end{minipage}
\begin{minipage}{4.25cm}
\includegraphics[width=4.25cm,clip]{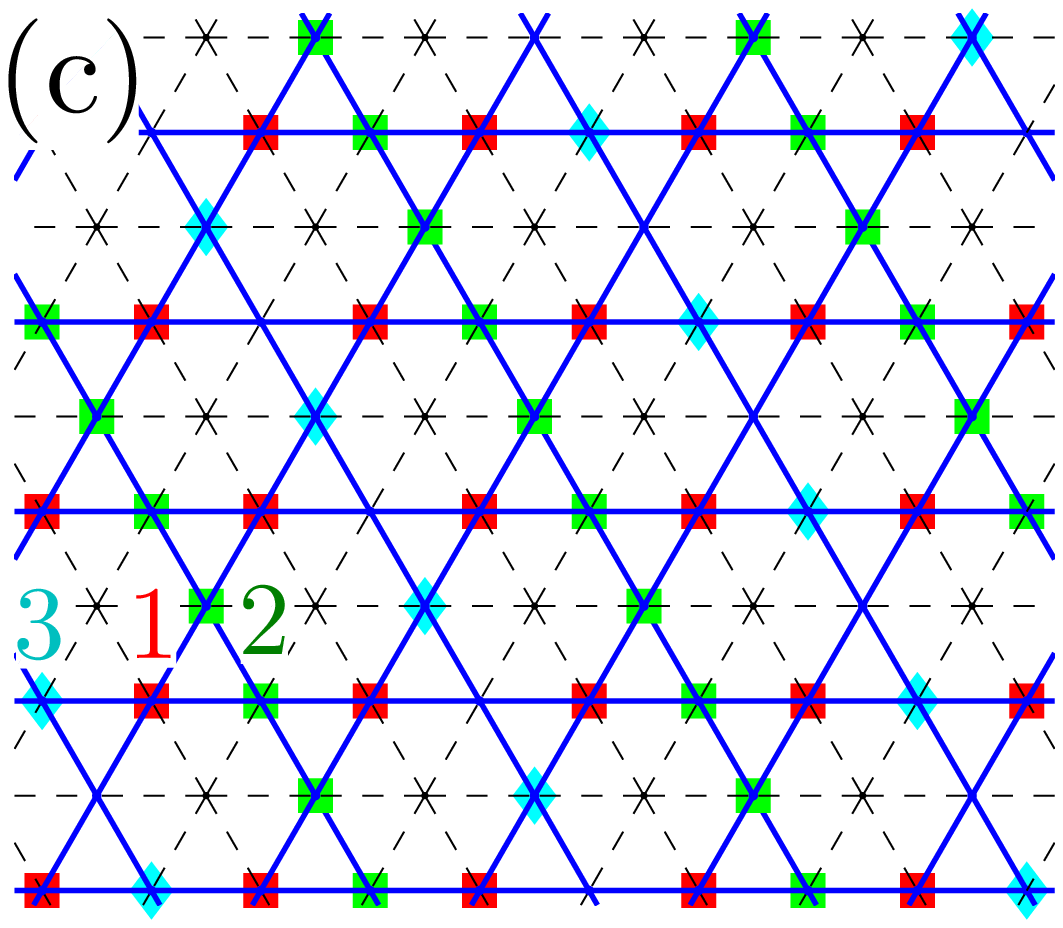}
\end{minipage}
\begin{minipage}{4.25cm}
\includegraphics[width=4.25cm,clip]{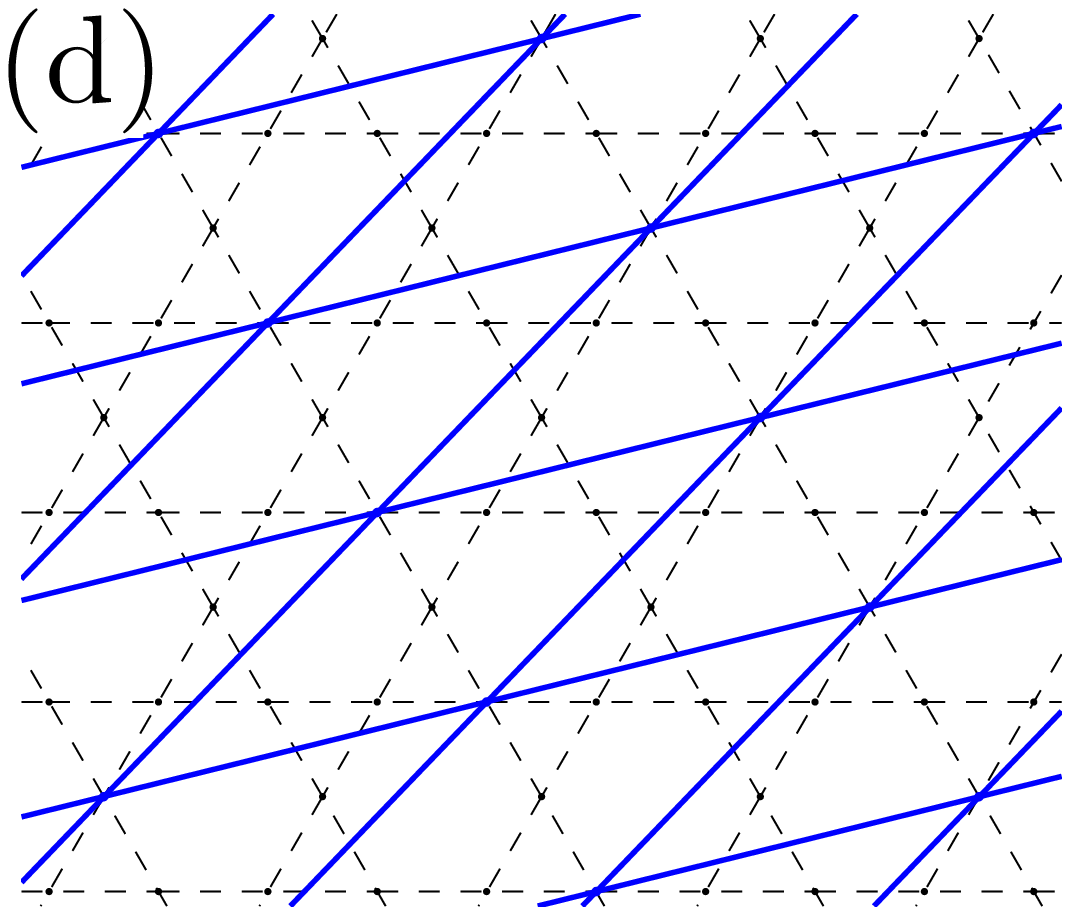}
\end{minipage}
\caption[]{\label{UniversalLattices} Graph states corresponding to (a) 
hexagonal, (b) triangular and (c) Kagome lattices are universal for MQC. LOCC transformation from (a) to  (d) (2D cluster)  via (b) and (c) is indicated, where simple graph rules can be used ($\sigma_y$ and $\sigma_z$ measurements are displayed by $\square$ and $\diamondsuit$, respectively).}
\end{figure}

{\bf Examples of universal resources.---} We now turn to the second main part of our analysis, where we obtain examples of families of states which \emph{are} universal resources. In particular, we show
 
{\bf Theorem 2:} \emph{The graph states corresponding to the hexagonal, triangular and Kagome lattices are universal.}
 
The proof is given in Fig. \ref{UniversalLattices}. 
We remark that other universal resource states have been presented \cite{Ch05} 
based on non-uniform lattice structures (see also \cite{Ta06}), where
each gate in a universal set of unitary gates can be implemented by
local measurements on an elementary unit and these units are combined
(bottom-up approach).
Here we use a different approach, where we prove universality by explicitly constructing LOCC protocols that yield the 2D cluster state (top--down approach). The proofs are based on successive applications of simple rules how to update a graph when applying certain local operations on the corresponding graph state \cite{He06}, where we only use local complementation (inversion of neighborhood graph) and vertex deletion, corresponding to $\sigma_y$- and $\sigma_z$-measurements, respectively. Note that the square 2D lattice, together with the hexagonal and triangular lattices are the only possibilities to obtain a regular tiling of the 2D plane. Furthermore, the Kagome lattice is an example of a uniform semiregular 2D tiling with 2 basic tiles (the triangle and the hexagon). The hexagonal lattice has vertex degree 3, which leads to an increased robustness against local noise as compared to the 2D cluster state \cite{Du04}.

So far we have demanded that LOCC succeed with unit success probability in order to call a family of states universal. One may also define {\em weak universality} of a family of states by considering probabilistic transformations rather than deterministic ones. A family of states is called a weak universal resource if an arbitrary unitary transformation can be implemented with probability $p_{\epsilon}=1-\epsilon$ for any $\epsilon>0$. One can follow the same approach as in the deterministic case, where one has to replace LOCC by {\em stochastic} LOCC (SLOCC). However, some care is required, in particular when formulating necessary conditions for weak universality in terms of entanglement measures.
For instance, any family of states that allows one to implement an arbitrary unitary operations with some (arbitrary small) non--zero probability of success becomes weakly universal when considering an extended family containing many copies of each of the states, even though the value of certain entanglement measures (e.g. maximal localizable entanglement $E^L_{ab}$) is strictly smaller than for $k\times k$ cluster states.
A proper necessary condition can be formulated by considering SLOCC orbits of all states within a given family. Any entanglement measure whose maximum value on the orbits of all states of the family is smaller than for the $k_0\times k_0$ cluster state for some $k_0$, allows one to deduce that the family is not a weak universal resource. 

It is straightforward to obtain examples of non--graph states that are weak universal resources. Consider the family of states defined by deformed cluster states, $|\psi_{k\times k}\rangle \propto \Lambda^{\otimes k^2} |C_{k\times k}\rangle$ with $\Lambda=\diag(1,\lambda)$ and $\lambda <1$, i.e. these states can be probabilistically obtained from the 2D cluster state by applying local filtering operations. The inverse transformation is also possible, however the success probability is exponentially small. Nevertheless, a single copy of a state in this class is weakly universal when $\lambda$ lies above a certain threshold. To see this,  note that one can deterministically transform a state $|\psi_{k\times k}\rangle $ by means of LOCC into a graph state corresponding to a 2D lattice with defects, by applying local 2--outcome measurements described by $\{\Lambda^{-1}=\diag(\lambda,1),\ \overline{\Lambda^{-1}}=\diag(\sqrt{1-\lambda^2},0)\}$ at each qubit. One finds that the defect probability $p_{\mbox{\scriptsize def}}=(1-\lambda^2)/(1+\lambda^2)$ is independent of the system size, and for $\lambda > \lambda_0 \approx 0.98$ one can indeed show that the resulting resource is still weakly universal. This corresponds to a percolation--type effect, and one can expect weak universality also for smaller values of $\lambda$.

{\bf Conclusion and outlook. --- }  In this letter we have analyzed universality of resources for MQC and the role of entanglement in this context. We have shown that the  entanglement width must diverge on any universal resource, and that all three regular tilings of the 2D plane are universal. A more detailed discussion about weak universality, simulations with non--unit accuracy, as well as efficiency of the simulation, will be reported elsewhere. Finally, we note that there is a close connection between unboundedness of  entanglement width of graph states and undecidability of (monadic second--order) logic. This connection will also be published in forthcoming work.

M.V.D.N. acknowledges valuable discussions with R. Raussendorf. This work was supported by the FWF, the European Union (OLAQUI, SCALA, QICS), the \"OAW through project APART (W.D.), and JSPS (A.M.).
\bibliographystyle{unsrt}

\begin{thebibliography}{99}

\bibitem{Ra01}
R. Raussendorf and H. J. Briegel, Phys. Rev. Lett. {\bf 86}, 5188 (2001);
Quant. Inf. Comp. {\bf 2}(6), 443 (2002).

\bibitem{Go99}
D. Gottesman and I. Chuang, Nature {\bf 402}, 390 (1999). 

\bibitem{Ni03}
M. A. Nielsen, Phys. Lett. A {\bf 308}, 96 (2003);
D. Leung, Int. J. Quant. Info. {\bf 2}(1), 33 (2004).

\bibitem{Pe04}
S. Perdrix and P. Jorrand, quant-ph/0404146.

\bibitem{Jo03}
R. Jozsa and N. Linden, Proc. R. Soc. A {\bf 459}, 2011 (2003); 
G. Vidal, Phys. Rev. Lett. {\bf 91}, 147902 (2003).

\bibitem{Br01}
H. J. Briegel and R. Raussendorf, Phys. Rev. Lett. {\bf 86}, 910 (2001).

\bibitem{Ve04}
F. Verstraete and J.I. Cirac, Phys. Rev. A. {\bf 70}, 060302(R) (2004); 
P. Aliferis and D. W. Leung, Phys. Rev. A {\bf 70}, 062314 (2004); 
P. Jorrand and S. Perdrix, quant-ph/0404125.


\bibitem{Ch05}
A. M. Childs, D. W. Leung, and M. A. Nielsen, Phys. Rev. A {\bf 71}, 032318 (2005).

\bibitem{Ni05}

M.A. Nielsen, quant-ph/0504097; I. Markov and Y. Shi, quant-ph/0511069; 
R. Jozsa, quant-ph/0603163.

\bibitem{Gu04}
O. G\"uhne {\it et al.}, Phys. Rev. Lett. {\bf 95}, 120405 (2005).

\bibitem{He06}
M. Hein {\it et al.}, quant-ph/0602096.

\bibitem{Foot}
Here we disregard the efficiency of the preparation and/or classical description of the states in $\Psi$.


\bibitem{Oum}
S.-I. Oum, PhD thesis, Princeton University, 2005.

\bibitem{Go00} 
M.C. Golumbic and U. Rotics, Int. J. Found. Comput. Sci. {\bf 11}, 423 (2000).

\bibitem{dist_her}
The graph $C_k$ is \emph{distance--hereditary} and therefore its rank width cannot be greater than one \cite{Oum}.

\bibitem{twd}
The tree width is a graph invariant that measures how close a graph is to a tree. The tree width of a tree is 1.

\bibitem{cwd}
One has $\mbox{cwd} \leq 4\cdot 2^{\mbox{\scriptsize twd} -1}+1$ \cite{Co00} and $\mbox{rwd} \leq \mbox{cwd}$ \cite{Oum}.

\bibitem{Co00}
B. Courcelle and S. Olariu, Discr. Appl. Math. {\bf 101}(1-3), 77 (2000).

\bibitem{Ve_loc04}
F. Verstraete, M. Popp, and J.I. Cirac, Phys. Rev. Lett. {\bf 92}, 027901 (2004).

\bibitem{Go05}
G. Gour, D. A. Meyer, and B.C. Sanders, Phys. Rev. A {\bf 72}, 042329 (2005).

\bibitem{Ta06}
M.S. Tame {\it et al.}, Phys. Rev. A {\bf 73}, 022309 (2006).

\bibitem{Du04}
W. D\"ur and H. J. Briegel, Phys. Rev. Lett. {\bf 92}, 180403 (2004).

\end{thebibliography}

\end{document}